\documentclass[twocolumn,showpacs,prl,amsmath,amssymb]{revtex4}
\ifx\pdftexversion\undefined
  \usepackage[dvips]{graphicx}
\else
  \usepackage[pdftex]{graphicx}
\fi

\begin{document}
\title{Breakdown of weak-field magnetotransport 
at a metallic quantum critical point}

\author{J.~Fenton}
\affiliation{School of Physics and Astronomy, University of
Birmingham, Birmingham B15 2TT, United Kingdom.}
 
\author{A.~J.~Schofield}
\affiliation{School of Physics and Astronomy, University of
Birmingham, Birmingham B15 2TT, United Kingdom.}

\date{\today}

\begin{abstract}
We show how the collapse of an energy scale in a quantum critical
metal can lead to physics beyond the weak-field limit usually used to
compute transport quantities.  For a density-wave transition we show
that the presence of a finite magnetic field at the critical point
leads to discontinuities in the transport coefficients as temperature
tends to zero. The origin of these discontinuities lies in the
breakdown of the weak field Jones-Zener expansion which has previously
been used to argue that magneto-transport coefficients are continuous
at simple quantum critical points. The presence of potential
scattering and magnetic breakdown rounds the discontinuities over a
window determined by $\tau \Delta < 1$ where $\Delta$ is the order
parameter and $\tau$ is the quasiparticle elastic lifetime.
\end{abstract}

\pacs{75.10.Lp, 72.15.Eb, 71.10.Hf, 71.18.+y}
\maketitle

Quantum critical points in metallic systems have generated
considerable theoretical and experimental interest in recent
years~\cite{coleman_2005a}. They are realized by tuning the finite
temperature critical point of a phase transition in a metal to zero
temperature. A variety of tuning parameters have been used including
hydrostatic pressure~\cite{julian_1996a}, chemical
composition~\cite{vonlohneysen_1994a} and quantum criticality can occur
serendipitously~\cite{gegenwart_1999a}. In insulating quantum magnets,
magnetic fields often provide a "handle" by which they may be made
quantum critical~\cite{bitko_1996a} and there has recently
been much interest in doing the same in the metallic case. Notable
examples include the metamagnetic quantum critical endpoint seen in
${\rm Sr_3Ru_2O_7}$~\cite{grigera_2001a} and also the
antiferromagnetic quantum critical point in ${\rm
YbRh_2Si_2}$~\cite{gegenwart_2002a}. The collapse of the
characteristic energy scale for fluctuations can induce deviations
from Landau Fermi liquid theory~\cite{hertz_1976a,millis_1993a} seen,
for example, in transport quantities. These are usually computed by
assuming linear response to driving fields, yet the vanishing energy
scale could invalidate that assumption~\cite{green_2005b}.  

In this paper we explicitly demonstrate such a breakdown by
considering magneto-transport in the simplest class of quantum
critical metal tuned by magnetic field.  The presence of the small but
finite magnetic field at the transition point leads to modified
transport response. This has assumed added significance recently
because of the suggestion that the discrepancies between theory and
experiment in quantum critical matter~\cite{coleman_2005a} may
originate in part from a transition at the critical point between
localized and de-localized spins and with a concomitant change in the
Fermi surface volume~\cite{coleman_2001a}. It is suggested that the
Hall coefficient would reveal this volume change and recently
evidence~\cite{paschen_2004a} for such a change has been presented at
the field driven quantum critical point in ${\rm YbRh_2Si_2}$.  Here,
we consider the $T\rightarrow 0$ limit of a spin or charge-density wave
transition in finite magnetic field, so the quantum critical
fluctuations play a small role in quasi-particle scattering when
compared with elastic scattering from impurities. This limit has been
considered by Bazaliy {\em et al.}~\cite{norman_2003a,bazaliy_2004a}
recently for non-field driven quantum critical points under the
assumption that a weak-field expansion in magnetic field can be
made. They conclude that there are no anomalies in the Hall
conductivity and that, in the absence of perfect nesting, changes in
other transport quantities are generally linear in the energy gap so
change continuously at the critical point.

\begin{figure}
%%preprint
%\includegraphics[width=0.9\columnwidth]{summary}
%%twocolumn
\includegraphics[width=\columnwidth]{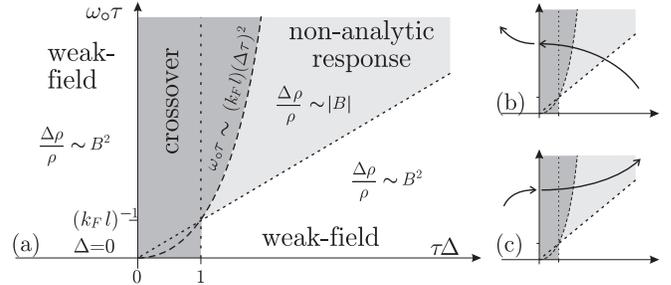}
\caption{\label{summary} (a) Summary of magnetotransport at a zero
temperature density wave (DW) 
transition as a function of the dimensionless gap,
$\tau \Delta$, and magnetic field, $\omega_\circ \tau$. In the
relaxation time approximation there is a discontinuity in transport
coefficients along the line $\tau \Delta =0$ due to the collapse of
the weak-field regime. Disorder and magnetic breakdown will
smooth this discontinuity into a crossover. A field-driven quantum
critical point is characterize by a trajectory through this phase diagram:
(b) a magnetic field suppressing the DW phase, or (b) stabilizing the
DW phase.}
\end{figure}

The essential result of our work is that at a density wave (DW)
quantum critical point the weak-field (Jones-Zener) expansion---upon
which previous magnetotransport work was based---breaks down because
of the vanishing DW gap $\Delta$.  This breakdown leads, in the
relaxation time approximation, to discontinuities in the
magnetoconductivities: jumps in the resistivity and Hall coefficients
at the transition, despite the fact that the phase transition itself
is continuous. For magnetic fields, $B > \Delta/(e v_F^2 \tau)$ we
show that the magnetoresistance is non-analytic in the field ($\sim
|B|$)---a result not obtainable from a weak-field expansion. While our
results are valid at any zero temperature DW transition, they assume
particular significance at a field driven quantum critical point where
$\Delta$ is tuned to zero at finite magnetic field. The
interdependence of the gap and magnetic field define a trajectory
through Fig.~\ref{summary}(a) [shown schematically in
Figs.~\ref{summary}(b) or (c)] which inevitably explores the
non-analytic region. We also go beyond the relaxation time
approximation to consider the effects of disorder and magnetic
breakdown very close to the critical point. While these smooth the
discontinuity they do not alter the magnitude of the changes in
conductivity across the transition at a finite magnetic field.

We consider the following highly simplified model of transport near a
density-wave transition (DW). We take the Fermi surface in the
paramagnetic state to be circular (and assume two dimensionality for
ease of computation). We then consider density wave formation to
induce a periodic potential on the otherwise free electrons and assume
this potential to be proportional to the mean-field order parameter:
$V_{\pm Q}(B)= \Delta_0 \Re{\sqrt{(B_c-B)/B_c}}$.
Here, the periodic potential (assumed real so both signs of $Q$ are
present) has Fourier components at the ordering wave-vector $Q$ of the
DW and its perturbation on the free electrons describes Bragg
scattering off the DW. Degenerate perturbation theory in the presence
of this potential will gap the dispersion whenever a resonance
condition is met: $\epsilon_{\vec{k}}=\epsilon_{\vec{k}\pm
n\vec{Q}}$. Since transport in the $T \rightarrow
0$ limit that we will be considering is dominated by the Fermi
surface, we need only consider the most important Bragg scattering
matrix element.  For example, near the point
$\epsilon_{k}=\epsilon_{k-Q}$ the dispersion is modified to be
\begin{equation}
E_k=\frac{1}{2}\left(\epsilon_k + \epsilon_{k-Q}\right) \pm
\sqrt{\frac{1}{4} \left( \epsilon_k - \epsilon_{k-Q} \right) +
V_{Q}^2} \; .
\label{hybridize}
\end{equation}
When the chemical potential falls in one of the gaps, as it will if
$2k_F>Q$, then the Fermi surface is modified by the density wave and
it is the change in transport properties induced by this modification
that we wish to study.

\begin{figure}
%%preprint
%\includegraphics[width=0.9\columnwidth]{sdw-FermiSurf}
%%twocolumn
\includegraphics[width=\columnwidth]{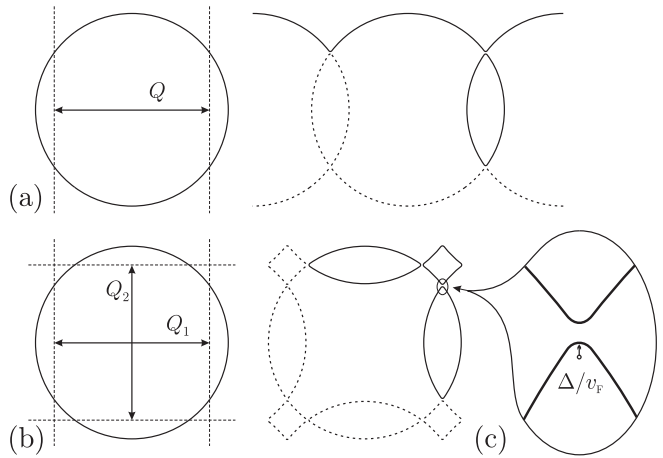}
\caption{\label{fermisurface} Fermi surface reconstructions at a
density wave transition. 
The solid lines indicate the Fermi surface to be used for
computing transport, while the dashed Fermi surfaces show the
dispersion in an extended zone scheme.
(a) An incommensurate DW transition
leads to open Fermi surface sheets. 
(b) At a commensurate DW transition 
($\vec{Q}_2 = \vec{Q}_1 + \vec{g}$)
the Fermi surface can break into closed pockets. 
The small gap near the DW
transition gives a local radius of curvature of the Fermi surface
$\sim \Delta/v_F$ which leads to a breakdown of the weak-field
expansion in the limit $\Delta \rightarrow 0^+$.}
\end{figure}

In Fig.~\ref{fermisurface} we illustrate the change in Fermi surface
topology that we envisage. There are a number of possible scenarios
\begin{itemize}
\item $2k_F<Q$: no gaps appear at the Fermi surface and transport
  coefficients will change smoothly through the critical point. We
  therefore do not consider this case here.
\item $2k_F>Q$, $Q$ incommensurate: gaps appear on the Fermi surface
  and lead to an open section of Fermi surface
  [Fig.~\ref{fermisurface}a]. In the absence of twinning, transport
  properties would discriminate between currents parallel and
  perpendicular to $\vec{Q}$.
\item $2k_F >Q$, $Q$ commensurate with the lattice ({\it e.g.} an
  antiferromagnet). Transport is isotropic through the transition and
  the Fermi surface remains closed.
  We treat this case by considering two real DWs with $\vec{Q}=(\pm
  Q_x, \pm Q_y)$ and $\vec{Q}=(\pm Q_x, \mp Q_y)$ (see
  Fig.~\ref{fermisurface}b).
\item $2k_F =Q$ (nested)~\cite{norman_2003a}. Since 
  the perfect nesting not be met very close to the critical 
  point~\cite{bazaliy_2004a} this case will ultimately reduce to one
  of the cases above and is therefore not considered here.
\end{itemize}

 For the initial analysis we use the classical Boltzmann equation in
the relaxation time approximation $\tau$: we are envisaging the $T=0$
limit of the conductivities and ignore inelastic processes. 
Rather than solve the Boltzmann equation order by order in the magnetic 
field~\cite{ziman_1960a} we solve to all orders in the field 
directly~\cite{abrikosov_1988a} using the Chamber's
formula~\cite{chambers_1969a}:
\begin{equation}
\sigma_{ij}=\frac{e^2}{4 \pi^3} \oint \frac{dS}{\hbar |\vec{v}|} \int_0^\infty v_i(0)v_j(t) e^{-t/\tau} dt \; .
\label{chambers}
\end{equation}
For each area element of the Fermi surface, $dS$, we integrate the
velocity, $\vec{v}(t)$ measured along a semiclassical quasiparticle
orbit. These orbits are defined by the Lorentz equation of motion
\begin{equation}
\hbar \frac{d\vec{k}}{dt}=-e \vec{v} \times \vec{B} \; ,
\label{lorentz}
\end{equation}
where $\vec{v}=\vec{\nabla}_k \epsilon(\vec{k})/\hbar$. In this paper
we will always assume the magnetic field is perpendicular to the 2D
electron fluid.

\begin{figure}
%%preprint
%\includegraphics[width=0.9\columnwidth]{sigma}
%%twocolumn
\includegraphics[width=\columnwidth]{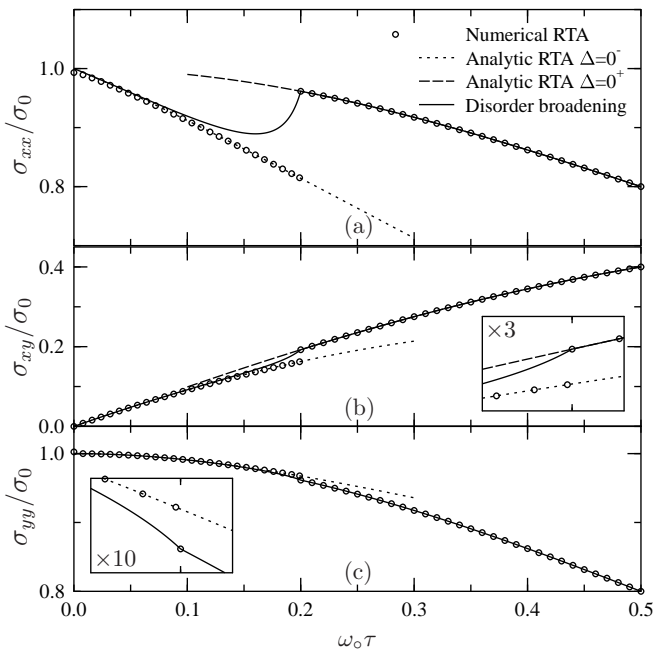}
\caption{\label{conductivities} The components of the conductivity
tensor (a) $\sigma_{xx}$, (b) $\sigma_{xy}$ and (c) $\sigma_{yy}$
(in units of $\sigma_0=ne^2 \tau/m$) computed when a mean-field
like DW gap is pushed to zero at a magnetic field equivalent to
$\omega_\circ \tau=0.2$. The circles are the numerical
solution of the Chamber's formula within the relaxation time
approximation (RTA).  The dashed lines show the analytic results. The solid
line goes beyond the relaxation time approximation by including the
role of disorder in washing out the DW gap.}
\end{figure}

Our main numerical results are illustrated in
Fig.~\ref{conductivities}.  The circles illustrate the conductivities
obtained by integrating the Chambers formula when an DW gap with $Q$
parallel to the $y$ axis, is suppressed to zero at a critical field
corresponding to $\omega_\circ \tau=0.2$ with
$\Delta_0/\epsilon_F=0.01$.  Here $\omega_\circ=eB/m$, the precession
frequency around the whole circular Fermi surface in the absence of a
DW and the transition is occurring in the region where $\omega_\circ
\tau \ll 1$---{\it i.e.} na\"\i vely the weak field regime.  We
clearly see discontinuities in the conductivities $\sigma_{xx}$ and
$\sigma_{xy}$ (and a very small one in $\sigma_{yy}$) at the critical
field.  In the case of closed Fermi surfaces (not shown) a clear
discontinuity occurs in all three conductivities. Having observed that
the conductivities can be discontinuous at a DW transition in a
magnetic field we now consider the underlying physics.

Within Boltzmann transport theory the effect of a magnetic field on
the conductivity is caused by the precession of the out-of-equilibrium
distribution around the Fermi surface according to Eq.~\ref{lorentz}
between scattering events. Quasi-particles are deflected by the local
Hall angle which is related to the local radius of curvature of the
Fermi surface ($R$) and the Lorentz equation of motion:
\begin{equation}
\delta k = R \theta_{\rm H} = e v_F B \tau \quad \Rightarrow \quad
\theta_{\rm H} = \frac{e v_F B}{R} \; .
\end{equation}
This deflects the current, leading to a reduction of the longitudinal
conductivity $\Delta \sigma_{xx}/\sigma_{xx}(0) \sim \cos \theta_H - 1
\sim \theta_H^2$ giving the usual magnetoconductance quadratic in
applied field. The Hall angle must be small for weak-field response.
However at a DW transition a gap opens in the Fermi surface leading
to a local radius of curvature of the Fermi surface of order
$\Delta/v_F$ which vanishes at the quantum critical point itself [see
Fig.~\ref{fermisurface}(c)]. The condition for being in the weak-field
regime is therefore that $B_{\rm weak-field} < \Delta/(e v_F^2 \tau)$
or equivalently $\omega_\circ \tau < \tau \Delta/(k_F l)$ where $l$ is
the mean free path.  This upper bound vanishes at the critical point
so magnetotransport there is never weak-field. Thus the results of
Bazaliy~{\it et al.} are valid only provided the limit $B/\Delta
\rightarrow 0$ can be taken which is certainly not the case at a
field-driven DW quantum critical point.

Instead magneto-transport is dominated by a fraction of the
quasiparticles (proportional to $e v_F B \tau$) that are deflected
around the cusp in the Fermi surface between scattering events.  This
leads to a magnetoconductance proportional to $|B|$---{\it i.e.}
non-analytic in the field and thus beyond any weak-field
expansion. 
This physics is beautifully illustrated by the
magnetoresistance of a square Fermi surface calculated by
Pippard~\cite{pippard_1989a}.  
%Note that the deflection primarily
%changes the component of velocity perpendicular to the DW $Q$ so only
%the longitudinal conductivity components perpendicular to $Q$ show a
%large discontinuity in conductance.

We can use Pippard's method to compute the size of the discontinuities
in all three independent components of the conductivity tensor. In the
limit that $\Delta \rightarrow 0^+$ we can solve the Boltzmann
transport equation on each segment of the Fermi surface and then
combine the solutions by ensuring that the magnitude of the
distribution function is continuous. The
result is that
\begin{eqnarray}
\label{pippxx}
\sigma_{xx}&=&\frac{ne^2\tau}{m}
\left[\frac{1}{1+(\omega_\circ \tau)^2} 
-
\frac{Q^2}{k^2_F}
\frac{\omega_\circ \tau P(\alpha,\omega_\circ \tau)}
{\left[1+(\omega_\circ \tau)^2\right]^2}
\right]\;, \\
\label{pippxy}
\sigma_{xy}&=&\omega_\circ \tau\sigma_{xx}\;, \\
\label{pippyy}
\sigma_{yy}&=&\frac{ne^2\tau}{m}
\left[\frac{1}{1+(\omega_\circ \tau)^2} 
+
\frac{Q^2}{k^2_F}
\frac{(\omega_\circ \tau)^3 P(\alpha,\omega_\circ \tau)}
{\left[1+(\omega_\circ \tau)^2\right]^2}
\right], 
\end{eqnarray}
where $\alpha=\cos^{-1} (Q/2k_F)$ and 
\begin{equation}
P(\alpha,\omega_\circ \tau)=\frac{\cosh\frac{\pi}{2\omega_\circ \tau}}
{\pi\cosh\frac{\alpha}{\omega_\circ \tau}
\sinh\frac{\pi/2-\alpha}{\omega_\circ \tau}}\;. 
\end{equation}
On the paramagnetic side ($\Delta \rightarrow 0^-$) the conductivities
are as above but without the terms proportional to $P(\alpha,
\omega_\circ \tau)$. Thus the all three magnetoconductivities,
$\sigma_{xx}$, $\sigma_{xy}$, $\sigma{yy}$, show discontinuities with
magnitude of order, $\omega_\circ \tau$, $(\omega_\circ \tau)^2$ and
$(\omega_\circ \tau)^3$ respectively so the most
dramatic jump is the magnetoresistance~{footnote}. These analytic solutions are
shown as the dashed lines in Fig.~\ref{conductivities}.
  
We now extend our calculation beyond the relaxation time
approximation. If $\tau \Delta<1$ this calculation cannot be valid
since the quasiparticles would be unable to notice the Bragg
scattering from the DW above the scattering from impurities. One would
expect this to washout the discontinuities in magnetoconductivities
since quasiparticles will tend to remain on original Fermi
surface. Magnetic (Zener) breakdown when $B > 2 \Delta^2/(e
v_F^2)$~\cite{abrikosov_1988a} as the same effect.  (See Green and
Sondhi~\cite{green_2005b} for $E$ field breakdown.)  We include this
phenomenologically in our calculation in the following fashion. Rather
than hybridize the dispersion (as in Eq.~\ref{hybridize}) we maintain
a circular Fermi surface and treat the DW potential in the collision
integral as a resonant scatterer between points of the Fermi surface
that satisfy the resonance condition. This gives the following
transport equation for the out of equilibrium distribution function
$g(\epsilon, \theta)$
%\begin{widetext}
\begin{eqnarray}
-e \vec{v} \cdot \vec{E} \tau \partial_\epsilon f_0 + \omega_\circ \tau
\frac{\partial g}{\partial \theta} = - g
-(\tau \Delta)^2 \left[\delta(\cos\theta+\eta) \right. \nonumber \\
  \left. +\delta(\cos\theta-\eta)\right]
\left[g(\epsilon,\theta)-g(\epsilon,\pi-\theta)\right] \; ,
\end{eqnarray}
%\end{widetext}
where $f_0$ is the Fermi function and $\eta=Q/2k$. Since the solution
is periodic around the Fermi surface it may be solved by Fourier
transform and, for a circular Fermi surface gives the following
expression for the conductivities
\begin{eqnarray}
\sigma_{xx}
\label{braggxx}
&=&\frac{ne^2\tau}{m}\left[\frac{1}{1+(\omega_\circ \tau)^2}
-\frac{K(\alpha,\omega_\circ \tau)}{(1+(\omega_\circ \tau)^2)^2}
\right]\;, \\
\label{braggxy}
\sigma_{xy}&=&\omega_\circ \tau\sigma_{xx}\;, \\
\label{braggyy}
\sigma_{yy}&=&\frac{ne^2\tau}{m}
\left[\frac{1}{1+(\omega_\circ \tau)^2} +
\frac{(\omega_\circ \tau)^2K(\alpha,\omega_\circ \tau)}
{(1+(\omega_\circ \tau)^2)^2}\right],
\end{eqnarray}
where
\begin{equation}
K(\alpha,\omega_\circ \tau)= \frac{2A \omega_\circ\tau
P(\alpha,\omega_\circ \tau)\cos^2\alpha}{\omega_\circ\tau
P(\alpha,\omega_\circ \tau)+A}\; ,
\end{equation} 
and $ A= \frac{4(\tau \Delta)^2}{\pi\sin\alpha}$. These conductivities
are shown as solid lines on Figs.~\ref{conductivities}. 
Note how these expressions
interpolate between the paramagnetic solution when $\tau \Delta=0$ (no
DW scattering), and the Pippard result in the limit $\tau \Delta \gg
1$ where the angle dependent scattering effectively mimics the
reconstructed the Fermi surface. Thus we see that disorder washes out
the discontinuity over a region in field determined by $\tau \Delta
<1$.

We have also considered the case of closed Fermi surfaces. In that
case $\sigma_{xx}=\sigma_{yy}$ and both show a discontinuity of
fractional order $\omega_\circ \tau$. The other difference from the
case of open Fermi surfaces is that, as for all open Fermi surfaces,
$\sigma_{yy}$ remains finite in the high field limit: $\omega_\circ
\tau \gg 1$. This is the regime where Landau level quantization of the
closed Fermi surfaces would also become important and is not
considered here.

In summary, we have shown that at a simple density wave (DW)
quantum critical point the weak-field regime of magnetotransport
collapses to zero field with the size of the gap. At finite field in a
clean metal one would expect to see discontinuities in the
magnetoresistance of order of the magnitude of the Hall angle:
$\omega_\circ\tau$. This effect will be significant at a field driven
quantum critical point where by definition the field is finite at the
transition point. The case of  ${\rm YbRh_2Si_2}$
already shows features in the Hall effect which suggest that it falls
outside the class of DW quantum critical points. A prediction from
this work is that non-field driven DW quantum critical points should
should a low field cross-over to a transverse magnetoresistance that
is linear in the applied field. It would be interesting to look for
such an effect in the $Cr_{1-x}V_x$ system under
pressure~\cite{yeh_2002a} where we estimate that $\tau \Delta=1$ at
$(x-x_c)/x_c \sim 0.1$ and $\omega_\circ \tau \sim (k_F l)^{-1}$ at $B
\sim 1$T. Very recently we have learnt that ${\rm Ca_3Ru_2O_7}$ shows
exactly the linear magnetoresistance we predict and is argued to be a
small gap density wave system~\cite{kikugawa_2005a}.

We acknowledge useful discussions with A. Rosch, C. Hooley, Q. Si,
Y. Bazaliy and R. Ramazashvili. JF acknowledges the EPSRC for
financial support and AJS acknowledges the Royal Society and
Leverhulme Trust for financial support and the hospitality of the
KITP, Santa Barbara. This research was supported in part by the
National Science Foundation under Grant No.~PHY99-0794.


\begin{thebibliography}{10}
\bibitem{footnote}
For the open Fermi surface case~[Fig.~\ref{fermisurface}(a)] the discontunities
  in Hall angle and number cancel, but this is not the case for the closed
  Fermi surface~[Fig.~\ref{fermisurface}(b)] where an order $(\omega_\circ
  \tau)^2$ jump remains in the Hall response.
\end{thebibliography}
\end{document}